\documentclass[superscriptaddress]{revtex4}
\usepackage{amsmath,lscape,epsfig}

\def\ii{\'{\i}}
\def\beq{\begin{equation}}
\def\eeq{\end{equation}}
\def\beqa{\begin{eqnarray}}
\def\eeqa{\end{eqnarray}}
\def\ban{\begin{eqnarray*}}
\def\ean{\end{eqnarray*}}
\def\bi{\begin{itemize}}
\def\ei{\end{itemize}}

\def\d{\mbox{d}}

\begin{document}

\title{Instabilities in asymmetric nuclear matter}

\author{S.S. Avancini}
\affiliation{Depto de F\'{\i}sica - CFM - Universidade Federal de Santa 
Catarina - Florian\'opolis - SC - CP. 476 - CEP 88.040 - 900 - Brazil}
\author{L. Brito}
\affiliation{Centro de F\ii sica Te\'orica - Depto de F\ii sica - 
Universidade de Coimbra - 3000 - Portugal}
\author{D.P.Menezes}
\affiliation{Depto de F\'{\i}sica - CFM - Universidade Federal de Santa 
Catarina - Florian\'opolis - SC - CP. 476 - CEP 88.040 - 900 - Brazil}
\author{C. Provid\^encia}
\affiliation{Centro de F\ii sica Te\'orica - Depto de F\ii sica - 
Universidade de Coimbra - 3000 - Portugal}

\begin{abstract}
The existence of phase transitions from liquid to gas phases in asymmetric 
nuclear matter (ANM) is related with the instability regions which are limited
by the spinodals. In this work we investigate the instabilities in ANM
described within relativistic mean field hadron models, both with constant 
and density dependent couplings at zero and finite temperatures.
In calculating the proton and neutron chemical potentials we have used an 
expansion in terms of Bessel functions that is convenient at low densities. 
The role of the isovector scalar $\delta$-meson is also investigated in the 
framework of relativistic mean field models  and density
dependent hadronic models. It is shown that the main differences occur at 
finite temperature and large isospin asymmetry close to the boundary of the
instability regions.  
\end{abstract}

\maketitle

\vspace{0.50cm}
PACS number(s): 21.65.+f, 21.90.+f, 24.10.Jv, 21.30.Fe

\section{Introduction}

Recently there has been a big development in the description of nuclei
and nuclear matter in terms of relativistic many body theory. In
particular, the phenomenological models developed using the
relativistic mean field theory describe well the ground state of both
stable and unstable nuclei \cite{sed,toki94}. These same models, with
conveniently adjusted parameters, are used to 
describe the properties of neutron stars and supernovae \cite{toki98,hp01}. 
Therefore, it is important to test these models  at finite temperature and 
different densities.  

The discussion of the properties of asymmetric nuclear matter (ANM) systems, 
namely their instabilities and phase transitions, are presently a topic 
of great insterest \cite{ms,bao02,bctl98,bctg01,lm01,mc03,lm03}. The 
instabilities present in ANM may manifest themselves as an isospin 
distillation or fractionation \cite{xu00}. 
It has recently been discussed
\cite{bctg01,mc03} that in an ANM system not only the
mechanical but also the chemical instabilities appear as an instability of 
the system against isoscalar fluctuations. Hence the spinodal
instability is dominated by density fluctuations 
which lead to a liquid gas separation with restoration of the isospin symmetry
in the dense phase.  This is known as the fractionation effect. 
Multifragmentation also
takes place when the system enters the spinodal region through nucleation or 
through spinodal decomposition. The correlation between spinodal 
decomposition and negative heat capacity evidences the fact that the spinodal 
decomposition is the dynamics underlying the liquid gas phase transition 
\cite{exp}. In the afore mentioned work, the authors point out the fact that 
systematic measurements including correlations between these signals and also 
the inclusion of the isospin degree of freedom are important sources of future
 experimental work.

On the other hand it is expected that neutron stars have a solid crust 
formed by non uniform neutron rich matter in $\beta$-equilibrium above a 
liquid mantle. In the inner crust, nuclei which form a lattice to reduce the
Coulomb energy, coexist with a gas of neutrons which have dripped out. The 
solid crust has an important role in explaining the sudden spin jumps known as
{\em glitches} in neutron stars \cite{glit}.  Recently it has been proposed 
that there is a relationship between the neutron skin of heavy nuclei and the 
properties of neutron star crusts \cite{hp01}, namely the thicker the 
neutron skin of a heavy nucleus the thinner the solid crust of a neutron star. 
Properties of the crust, namely its thickness and  pressure at the crust-core 
interface, depend largely on the density dependence of the equation of state 
(EOS). It is of particular importance the transition density below which 
uniform neutron rich  matter becomes unstable.  In  \cite{hp01} it was also 
shown that the thicker the neutron skin the lower the transition density form 
a uniform to a non uniform neutron rich matter.

Within the framework of relativistic models, the liquid gas phase transition 
in nuclear matter has been investigated at zero and finite temperatures for 
symmetric and asymmetric infinite systems \cite{ms,rs,md,epsw}.
With the help of the Thomas - Fermi approximation, we have investigated 
droplet formation in  the liquid gas phase 
transition in cold~\cite{mp1} and hot~\cite{mp2} asymmetric nuclear 
matters using relativistic mean field (RMF) models ~\cite{sed,bb}. 
In \cite{calorica} we have considered  a droplet immersed in a gas of
evaporated particles, in such a way that they mimic a source of changing
mass. As temperature increases  particles evaporate, mainly neutrons,  and 
the fraction of protons in the droplet increases leading to isospin 
fractionation. In \cite{pmb02} we have shown  that at finite temperature 
droplet properties within different parametrizations of the RMF model have 
different behaviors with temperature. 

The EOS of neutron rich matter is in particular sensitive to the density
dependence of the symmetry energy. In \cite{hp01} the effect of changing this 
dependence was studied through the inclusion of non linear $\sigma-\rho$ and
$\omega-\rho$ couplings. On the other hand the authors of \cite{lgbct} claim 
that the isovector scalar meson $\delta$ is of vital importance in finding 
the stability conditions of drip-line exotic nuclei because the structure of 
relativistic interactions with a scalar (attractive) and a vector (repulsive)
potential, which balance each other, is also present in the isovector
channel.  The $\delta$-field has also important effects on the
symmetry properties of the nuclear system. 

Standard RMF interactions have their limitations even for describing nuclei 
close to the stablility line. This is due to the fact that the isovector 
channel is poorly  constrained by experimental data. An example is the 
systematic overestimate of the neutron skins \cite{brown}. Some of these 
limitations are overcome by quantum hadrodynamical models with density 
dependent meson-nucleon couplings (which we refer to as TW) \cite{TW,ring02,
abcd03}, which have been used with success to describe both nuclear matter 
and finite nuclei. In these models the couplings are either taken from 
Dirac-Brueckner-Hartree-Fock calculations or are fitted to data of nuclear 
matter or finite nuclei.

We may ask whether the recent improvements of the RMF models, both  through
the  inclusion of  density dependent meson-nucleon couplings and/or the 
$\delta$ scalar-isovector meson, present different features at subnuclear 
densities of nuclear asymmetric matter  which could have consequences for the
properties of the inner crust of neutron stars or in multifragmentation or 
isospin fractionation reactions. The parametrizations of these models take 
generally into account saturation properties of nuclear matter and properties 
of stable nuclei. Extension of the model for very asymmetric nuclear matter or
to finite temperatures may show different behaviors. The simplest test 
between the models is a comparison of the regions of uniform unstable matter.

In the present work we direct out investigation to some of the topics
we have briefly mentioned before. We study the liquid gas phase transition 
and, in 
particular, determine the instability regions occuring within the density 
dependent hadronic models \cite{TW,gaitanos} and the non linear relativistic 
mean field models. We also  study models including the $\delta$  meson.

In order to determine the instability regions, we found convenient to 
generalize the  calculation of  the chemical potentials as prescribed in 
\cite{jaq84} to relativisitc models with density dependent effective
masses. The expansion we present in our paper is accurate in the range of 
temperatures and densities discussed  and allows us to perform explicit 
derivatives on the chemical potentials.

This paper is organized as follows: in section 2 we present the whole
formalism in which our calculations are based, namely the relativistic models,
the chemical potential expansions and the thermodynamical conditions for the
description of the instability regions; in section 3 a brief explanation 
regarding the introduction of the isovector-scalar meson is reported; in 
section 4 the results are presented and the conclusions are drawn. Finally we
include an appendix with some important formulae not given in the main 
text. 
 
\section{The formalism}

We start from the lagrangian density of the
relativistic TW model \cite{TW}
$$
{\cal L}=\bar \psi\left[\gamma_\mu\left(i\partial^{\mu}-\Gamma_v V^{\mu}-
\frac{\Gamma_{\rho}}{2}  \vec{\tau} \cdot \vec{b}^\mu \right) 
-(M-\Gamma_s \phi)\right]\psi
$$
\begin{equation}
+\frac{1}{2}(\partial_{\mu}\phi\partial^{\mu}\phi
-m_s^2 \phi^2) 
-\frac{1}{4}\Omega_{\mu\nu}\Omega^{\mu\nu}+\frac{1}{2}
m_v^2 V_{\mu}V^{\mu}
-\frac{1}{4}\vec B_{\mu\nu}\cdot\vec B^{\mu\nu}+\frac{1}{2}
m_\rho^2 \vec b_{\mu}\cdot \vec b^{\mu}  ,
\label{lagtw}
\end{equation}
where
$\Omega_{\mu\nu}=\partial_{\mu}V_{\nu}-\partial_{\nu}V_{\mu}$ ,
$\vec B_{\mu\nu}=\partial_{\mu}\vec b_{\nu}-\partial_{\nu} \vec b_{\mu}
- \Gamma_\rho (\vec b_\mu \times \vec b_\nu)$.
The  parameters of the model are:
the nucleon mass $M=939$ MeV, the masses of
the mesons $m_s$, $m_v$, $m_\rho$, 
and the density dependent coupling constants $\Gamma_{s}$, 
$\Gamma_v$ and $\Gamma_{\rho}$, which are adjusted in order to reproduce
some of the nuclear matter bulk properties, using the following 
parametrization:
\begin{equation}
\Gamma_i(\rho)=\Gamma_i(\rho_{sat})f_i(x), \quad i=s,v
\label{paratw1}
\end{equation}
with
\begin{equation}
f_i(x)=a_i \frac{1+b_i(x+d_i)^2}{1+c_i(x+d_i)^2},
\end{equation}
where $x=\rho/\rho_{sat}$ and
\begin{equation}
\Gamma_{\rho}(\rho)=\Gamma_{\rho}(\rho_{sat}) \exp[-a_{\rho}(x-1)],
\label{paratw2}
\end{equation}
with the values of the parameters $m_i$, $\Gamma_i$, $a_i$, $b_i$, $c_i$ and 
$d_i$, $i=s,v,\rho$  given in \cite{TW}. Other possibilities
for these parameters are also found in the literature \cite{ditoro}.
Notice that in this model the non-linear terms are not present, in contrast 
with the usual non-linear  Walecka model  (NLWM).

Within the Thomas-Fermi approximation, the thermodynamic potential
is obtained. After it is minimized with respect to the  meson
fields, the following  equations are obtained
\begin{equation}
\phi_0= \frac{\Gamma_s}{m_s^2} \rho_{s},
\label{phi0}
\end{equation}
\begin{equation}
V_0 = \frac{\Gamma_v }{m_v^2} \rho,
\end{equation}
\begin{equation}
b_0 = \frac{\Gamma_{\rho}}{2 m_{\rho}^2} \rho_3, 
\label{b0}
\end{equation}
with
\begin{equation}
\rho=\rho_p+\rho_n,\quad \rho_3=\rho_p-\rho_n, \quad
\rho_i=2 \int\frac{\d^3p}{(2\pi)^3}(f_{i+}-f_{i-}), \qquad i=p,n 
\label{rhob}
\end{equation}
$$\rho_{s}=\rho_{s p}+\rho_{s n}, \qquad
\rho_{s i}=2 \int\frac{\d^3p}{(2\pi)^3} \frac{M^*}{E^{\ast}} (f_{i+}+f_{i-}),
$$
$M^*=M - \Gamma_{s}~ \phi_0$,  
$E^{\ast}=\sqrt{{\mathbf p}^2+{M^*}^2}$ and
$
f_{i\pm}=
{1}/\{1+\exp[(E^{\ast} \mp \nu_i)/T]\}\;$,
where the effective chemical potential is 
\begin{equation}
\nu_i=\mu_i - \Gamma_{v} V_0 -\tau_{i3} \frac{\Gamma_{\rho}}{2} b_0 - \Sigma_0^R,
\quad \tau_{p3}=1, \, \, \tau_{n3}=-1
\label{nu}
\end{equation}
with the rearrangement term given by 
\begin{equation}
\Sigma_0^R=\frac{\partial\, \Gamma_v}{\partial \rho}\, \rho\, V_0+\frac{\partial\, \Gamma_\rho}{\partial \rho}\, \rho_3\, \frac{b_0}{2}-
\frac{\partial\, \Gamma_s}{\partial \rho}\, \rho_s\, \phi_0.
\label{rearr}
\end{equation}
The energy density  in the mean field approximation reads:
\begin{eqnarray}
{\cal E}(\Gamma_s, \Gamma_v, \Gamma_\rho)&=
& 2 \sum_i \int \frac{\d^3p}{(2\pi)^3} E^{\ast} \left(f_{i+}+f_{i-}\right)+
\nonumber \\
&&\frac{m_s^2}{2} \phi_0^2 
+\frac{m_v^2}{2} V_0^2  
+\frac{m_{\rho}^2}{2} b_0^2.
\label{ener}
\end{eqnarray}
The pressure becomes
\begin{eqnarray}
P(\Gamma_s, \Gamma_v, \Gamma_\rho)&=&\frac{1}{3 \pi^2} \sum_{i}
\int \d p \frac{{\mathbf p}^4}{E^{\ast}}
\left( f_{i+} + f_{i-}\right)
-\frac{m_s^2}{2} \phi_0^2\left(1+2\frac{\rho}{\Gamma_s}\frac{\partial \, \Gamma_s}{\partial\rho}\right)
\nonumber \\
&+&\frac{m_v^2}{2} V_0^2\left(1+2\frac{\rho}{\Gamma_v}\frac{\partial \, \Gamma_v}{\partial\rho}\right) 
+\frac{m_{\rho}^2}{2} b_0^2 \left(1+2\frac{\rho}{\Gamma_\rho}\frac{\partial \, \Gamma_\rho}{\partial\rho}\right)
\label{press}
\end{eqnarray}

For two of the usual NLWM parametrizations,  namely NL3 \cite{nl3} and TM1 
\cite{tm1}, the above equations read: 
\begin{equation}
{\cal E}_{NL3}={\cal E}(g_s, g_v, g_\rho) +
\frac{\kappa\phi_0^3}{6} +\frac{\lambda\phi_0^4}{24},~~~
P_{NL3}=P(g_s, g_v, g_\rho) -\frac{\kappa\phi_0^3}{6} -
\frac{\lambda\phi_0^4}{24}
\label{epnl3}
\end{equation}
and
\begin{equation}
{\cal E}_{TM1}={\cal E}(g_s, g_v, g_\rho) +
\frac{\kappa\phi_0^3}{6} +\frac{\lambda\phi_0^4}{24}
+\frac{\xi g_v^4 V_0^4}{8},~
P_{TM1}=P(g_s,g_v,g_\rho) -\frac{\kappa\phi_0^3}{6} -
\frac{\lambda\phi_0^4}{24}+\frac{\xi g_v^4 V_0^4}{24},
\label{eptm1}
\end{equation}
where the meson-nucleon coupling constants, $g_s, g_v,$ and $g_\rho$ 
substitute $\Gamma_s$,$\Gamma_v$, and $\Gamma_\rho$. They  
are not density dependent and consequently all derivative terms in the 
pressure cancel out and 
$\kappa$, $\lambda$ and $\xi$ are the self-coupling constants multiplying 
the non-linear terms.

In order to study the instability region at low densities it is convenient to 
invert eq.(\ref{rhob}) and  obtain the effective chemical potential
that appear inside the distribution functions. We have followed the 
pres\-cription given in \cite{bao02,jaq84}, where just the 
particle distribution function is considered. The effective chemical potential
reads:
\begin{equation}
\nu_i=\frac{1}{\beta}\left(\ln(\eta_i)+\sum_{l=1}^\infty b_l \eta_i^l\right), 
\quad i=p,n 
\label {nuq}
\end{equation}
where
\begin{equation}
\eta_i=\frac{\rho_i}{ \gamma\, Q(\beta\, M^*)},
\end{equation}
$\gamma$ is the spin 
multiplicity, $\beta=1/T$ and
$$Q(\beta\, M^*)= \int\frac{d^3p}{(2\pi)^3} e^{-\beta E^*}=
\frac{{M^*}^2}{2\pi^2\, \beta}
K_2(\beta M^*),$$
with $K_n$ the modified Bessel function. The coefficients $b_l$ are defined in 
terms of the ratios
$$S_n=\frac{Q(n\beta\, M^*)}{Q(\beta\, M^*)}=\frac{K_2(n\beta\, M^*)}
{n\, K_2(\beta\, M^*)},$$
 and have been calculated in \cite{jaq84}. We list the first three
$$b_1=a_2, \quad b_2=a_3-a_2^2/2,\quad b_3=a_4-a_2 a_3+ a_2^3/3,$$
with 
$a_2=S_2(\beta\, M^*),$ $a_3=2S^2_2(\beta\, M^*)-S_3(\beta\, M^*),$
$a_4=5S^3_2(\beta\, M^*)-5\,S_3(\beta\, M^*)S_2(\beta\, M^*)+S_4(\beta\, M^*).$
The chemical potentials $\mu_i,\, i=p,n$  are obtained from equation
(\ref{nu}) and they read
\begin{equation}
\mu_i=\Gamma_v\, V_0+\tau_{i3}\frac{\Gamma_\rho}{2}b_0+ \Sigma_0^R+
\frac{1}{\beta}\left(\ln(\eta_i)+\sum_{l=1}^\infty b_l \eta_i^l\right)
\end{equation}
where the rearrangement term is given in equation (\ref{rearr}).

\begin{figure}
\begin{center}
\includegraphics[width=12.0cm,angle=0]{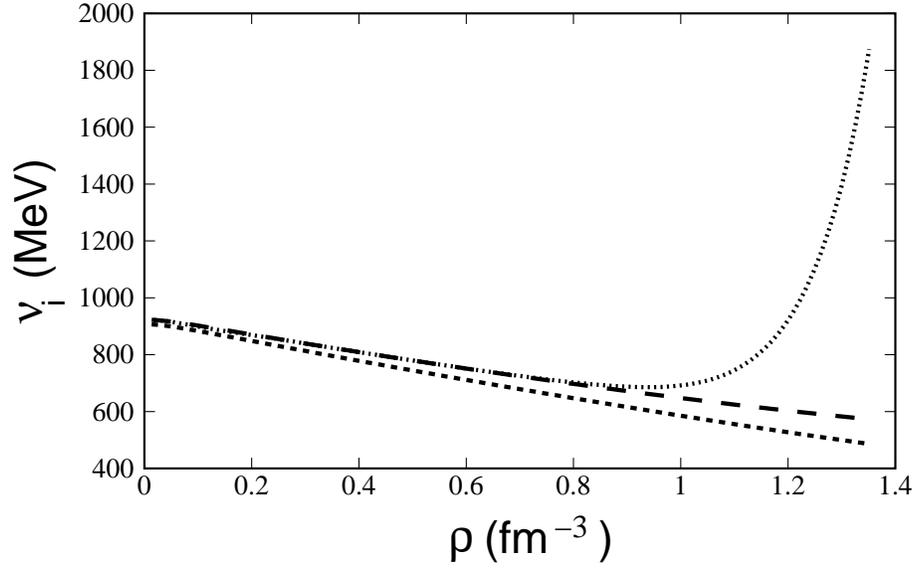}
\end{center}
\caption{Chemical potential curves for $T=7$ MeV, NL3 parameter set and
$y_p=0.1$. The curves for the exact proton chemical potential and its 
expansion are coincident (short dashed line). The straight long dashed line
represents the exact neutron chemical potential and the bending one (dotted
line) its expansion. The expansions are taken up to $5^{th}$ order.}
\label{fig1}
\end{figure}

We have checked the range of applicability of the expansion given in 
eq.(\ref{nuq}) by comparing it with the exact results. We have 
concluded that, for symmetric matter, the expansion works very well for 
temperatures $T \geq 7$ MeV and subsaturation densities, $\rho<\rho_0$. This 
agreement improves for higher 
temperatures within the range of temperatures involved in the present work, 
where the antiparticles do not play a crucial role. For ANM, $T=$7
MeV,  the agreement is still good for the proton chemical potential. The
neutron chemical potential is well reproduced only for
$\rho<0.8\rho_0$ as can  be seen in  fig. \ref{fig1}, where we have plotted 
the effective chemical 
potentials for protons and neutrons for asymmetric matter with the
proton fraction $y_p=0.1$, where $y_{ p}=\frac{\rho_{ p}}{\rho}$.
This result is still adequate
for the  range of densities we have considered in the present work

The stability conditions for asymmetric nuclear matter, keeping volume and 
temperature constant, are obtained from the free energy density $\cal F$, 
imposing that this function is a
convex function of the densities $\rho_p$ and $\rho_n$, i.e. the symmetric 
matrix with the elements \cite{ms,bctl98,mc03}
$${\cal F}_{ij}=\left(\frac{\partial^2{\cal F}}{\partial \rho_i\partial\rho_j}
\right)_T,$$
is positive.  
This is equivalent to imposing  \cite{landau}
\begin{equation}\frac{\partial \mu_p}{\partial\rho_p}>0,
\label{stab1}\end{equation}
\begin{equation}
\frac{\partial(\mu_p,\mu_n)}{\partial(\rho_p,\rho_n)}>0, \label{stab2}
\end{equation}
where we have used $\mu_i=\left.\frac{\partial{\cal F}}{\partial \rho_i}
\right|_{T,\rho_{j\ne i}}$.
In terms of  the proton fraction the
conditions (\ref{stab1}) and (\ref{stab2}) can be rewritten, respectively,  
in the form
$$\left(\frac{\partial P}{\partial \rho}\right)_{T,y_{p}}>0$$
and
\begin{equation}
\left(\frac{\partial P}{\partial \rho}\right)_{T,y_{p}}\,
\left(\frac{\partial \mu_p}{\partial y_{p}}\right)_{T,P}>0.
\label{spinodal}
\end{equation}
It has recently been argued \cite{bctg01,mc03} that in ANM the spinodal 
instabilities 
cannot be separately classified as mechanical or chemical instabilities. In 
fact, the two conditions that give rise to the instability of the system
are coupled so that it appears as a mixture of baryon density and
concentration fluctuations.
Therefore, we define the stability region as determined by (\ref{spinodal}).

In order to calculate the boundaries of the spinodal instability  regions we 
use the Gibbs-Duhem relation, at a fixed temperature and isospin asymmetry,
\begin{equation}
\left(\frac{\partial P}{\partial \rho}\right)_{T,\delta}=\frac{\rho}{2}\left
[(1+\delta)\frac{\partial \mu_n}{\partial
    \rho}+(1-\delta)\frac{\partial \mu_p}{\partial
    \rho}\right] = 
\sum_{i=p,n}
\rho_i \left(\frac{\partial \mu_i}{\partial \rho}\right)_\delta
\label{gb}
\end{equation}
with $\delta=-\rho_3/\rho=1-2y_p$, and
\begin{equation}
\left(\frac{\partial \mu}{\partial \delta}\right)_{T,P}=\left(\frac
{\partial \mu}{\partial
    \delta}\right)_{T,\rho}-\left(\frac{\partial \mu}{\partial
    \rho}\right)_{T,\delta}\, \left(\frac{\partial P}{\partial \rho}\right)^
{-1}_{T,\delta}\, 
\left(\frac{\partial P}{\partial \delta}\right)_{T,\rho},
\label{dmu}
\end{equation}
where
\begin{equation}
\left(\frac{\partial P}{\partial \delta}\right)_{T,\rho}=\frac{\rho}{2}
\left[(1+\delta)\frac{\partial \mu_n}{\partial \delta}+(1-\delta)\frac{
\partial \mu_p}{\partial\delta}\right]=
\sum_{i=p,n}
\rho_i \left(\frac{\partial \mu_i}{\partial \delta}\right)_\rho.
\label{outra}
\end{equation}
With the expressions given in this section the spinodal regions can be obtained
for both different temperatures and different  parametrizations.

\section{Including isovector-scalar mesons}

To investigate the influence of the $\delta$-meson in the stability conditions 
we have included in the NLWM the isovector-scalar meson terms 
\cite{lgbct}:
$$
{\cal L}=\bar \psi\left[\gamma_\mu\left(i\partial^{\mu}-g_v V^{\mu}-
\frac{g_{\rho}}{2}  \vec{\tau} \cdot \vec{b}^\mu \right) 
-(M-g_s \phi- g_{\delta} \vec{\tau} \cdot \vec{\delta})\right]\psi
$$
$$
+\frac{1}{2} \left(\partial_{\mu}\phi\partial^{\mu}\phi
-m_s^2 \phi^2 - \frac{1}{3}\kappa \phi^3-\frac{1}{12} \lambda \phi^4 \right) 
-\frac{1}{4}\Omega_{\mu\nu}\Omega^{\mu\nu}+\frac{1}{2}
m_v^2 V_{\mu}V^{\mu}
$$
\begin{equation}
-\frac{1}{4}\vec B_{\mu\nu}\cdot\vec B^{\mu\nu}+\frac{1}{2}
m_\rho^2 \vec b_{\mu}\cdot \vec b^{\mu} +\frac{1}{2}(\partial_{\mu} \vec 
\delta \partial^{\mu}\vec \delta
-m_{\delta}^2 {\vec \delta}^2\,), 
\label{lagdelta}
\end{equation}
where
$g_{\delta}$ and $m_{\delta}$ are respectively the coupling constant of the 
$\delta$ meson with the nucleons and its mass. Self-interacting terms for the 
$\sigma$-meson are also included, $\kappa$ and $\lambda$ denoting the 
corresponding coupling constants. The set of constants is defined by 
$g_i=\sqrt{f_i m_i^2}$, $i=s,v,\delta$, $\frac{g_\rho}{2}=\sqrt{f_\rho m_\rho^2}$,  $m_s=550$ MeV, $m_v=783$ MeV,
$m_{\rho}=763$ MeV, $m_{\delta}=980$ MeV,  
$f_s=10.33$ fm$^2$, $f_v=5.42$ fm$^2$, $f_{\rho}=3.15$ fm$^2$,
$f_{\delta}=2.5$ fm$^2$, $\kappa=0.066 g_s^3$ and 
$\lambda=-6 \times 0.0048 g_s^4$ \cite{lgbct} and we call it NL$\delta$.
For reference, in Table \ref{prop} we show the properties of nuclear matter reproduced 
by the models we discuss in the present work.
 From the minimization of the thermodynamic potential obtained in a 
Thomas-Fermi
approach, the equation of motion for this field becomes
\begin{equation}
\delta_3= \frac{g_{\delta}}{m_{\delta}^2} \rho_{s3},
\label{delta3}
\end{equation}
with $\rho_{s3}=\rho_{s p}-\rho_{s n}$. The energy density and the pressure 
are also affected by the presence of the new meson. The term 
$+ 1/2\, m_{\delta}^2 \,\delta_3^2$ should be added to the energy density
and the  $- 1/2 m_{\delta}^2 \delta_3^2$ should be added to the expression of 
the pressure, both given in eq. (\ref{epnl3}).
The effective masses for protons and neutrons acquire different values, namely,
$$
M^*_i=M - g_{s}~ \phi_0 -\tau_{i3} g_{\delta} \delta_3 \qquad i=p,n\,.
$$ 

\begin{figure}
\begin{center}
\includegraphics[width=12.0cm,angle=0]{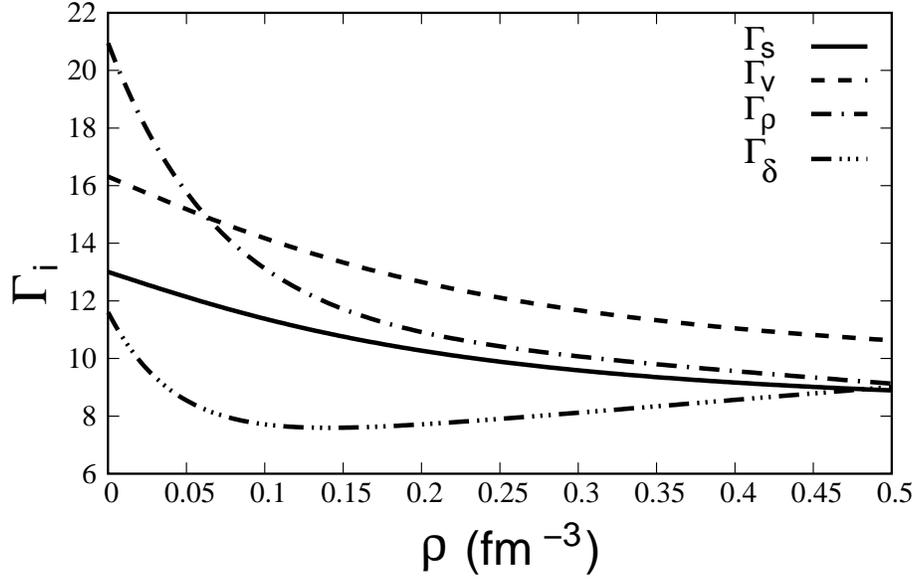}
\end{center}
\caption{Density dependent couplings as obtained from the proposed 
parametrization given in the text and which reproduce the curves given in
\cite{gaitanos}.}
\label{fig5}
\end{figure}

For completeness, we have also included the $\delta$-meson in a model where
the $\rho$ and $\delta$ couplings are density dependent, as done in 
\cite{gaitanos}, where it is called density dependent hadronic model 
(DDH$\rho \delta$). For this purpose, we have considered the density 
dependence of the
$\rho$ and $\delta$-nucleon vertices given in fig. 1 of ref. \cite{gaitanos} 
which have been extracted from DBHF calculations of ref. \cite{jong}.  
In this case the coupling constants $g_s$, $g_v$, $g_\rho$ and $g_\delta$
used in equation (\ref{lagdelta}) should be replaced by $\Gamma_s$, $\Gamma_v$,
$\Gamma_\rho$ and 
$\Gamma_\delta$ and the non-linear scalar terms do not appear.
For $\Gamma_s$ and $\Gamma_v$ we take the parametrizations given in
Eqs.(\ref{paratw1}) and (\ref{paratw2}). For $\Gamma_\rho$ and
$\Gamma_\delta$, we propose the following parametrization, 
$$\Gamma_i=\Gamma(\rho_{sat})\, f_i(x), \quad x=\rho/\rho_{sat}$$
with
$$f(x)=a_i \exp[-b_i(x-1)]-c_i(x-d_i), \qquad i=\rho,\,\delta.$$
and the parameters $a_i,b_i,c_i$ and $d_i$ are defined in Table \ref{para}.
This parametrization reproduces the curves 
given in fig. 1 of \cite{gaitanos} and is also displayed in fig. \ref{fig5}
of the present work.
We point out that in the present work there is a factor 2 difference
in the definition of $\rho$-meson coupling constant.
We take for the effective chemical potential Eq.~(\ref{nu}), with the
rearrangement term given by
$$
\Sigma_0^R=\frac{\partial\, \Gamma_v}{\partial \rho}\, \rho\, V_0+\frac{\partial\, \Gamma_\rho}{\partial \rho}\, \rho_3\, \frac{b_0}{2}-
\frac{\partial\, \Gamma_s}{\partial \rho}\, \rho_s\, \phi_0 -\frac{\partial\, \Gamma_\delta}{\partial \rho}\, \rho_{s3}\, \delta_3
$$
and for the pressure
$$
P(\Gamma_s, \Gamma_v, \Gamma_\rho, \Gamma_\delta)=P(\Gamma_s, \Gamma_v, \Gamma_\rho)
-\frac{m_\delta^2}{2} \delta_3^2\left(1+2\frac{\rho}{\Gamma_\delta}\frac{\partial
    \, \Gamma_\delta}{\partial\rho}\right).
$$

\section{Results and Conclusions}

A  quantity of interest in ANM is the nuclear bulk symmetry energy 
discussed in \cite{lgbct}. This quantity is important in studies involving 
neutron stars and radioactive nuclei. The behavior of the symmetry energy at 
densities larger than nuclear saturation density is still not well 
established. In general,
relativistic and non-relativistic models give different predictions for the
symmetry energy. 

It is usually defined as
\begin{equation}
{\cal E}_{sym} =\left. \frac{1}{2} \frac{\partial^2 {\cal E}/\rho}
{\partial \delta^2} \right|_{\delta=0},
\label{symen}
\end{equation}
which, for the models not including the $\delta$-meson, can be analytically 
rewritten as
\begin{equation}
{\cal E}_{sym}= \frac{k_F^2}{6 E^{*}_F}+ \frac{\Gamma_\rho^2}
{8 m_\rho^2} \rho, \label{esym}
\end{equation}
where
\[
k_{Fp}=k_F(1+\delta)^{1/3},\qquad k_{Fn}=k_F(1-\delta)^{1/3},
\]
with $k_F=(1.5 \pi^2\rho)^{1/3}$.
If the $\delta$ meson is included the symmetry energy reads \cite{lgbct}:
\begin{equation}
{\cal E}_{sym}= \frac{k_F^2}{6 E^*_F}+ \frac{\rho}{2} 
\left[  \frac{\Gamma_\rho^2}
{4 m_\rho^2} -  \frac{\Gamma_\delta^2}
{m_\delta^2} \left( \frac{M^*}{E^*_F} \right)^2 \right].
\label{esymd}
\end{equation}

\begin{figure}
\begin{center}
\includegraphics[width=12.cm,angle=0]{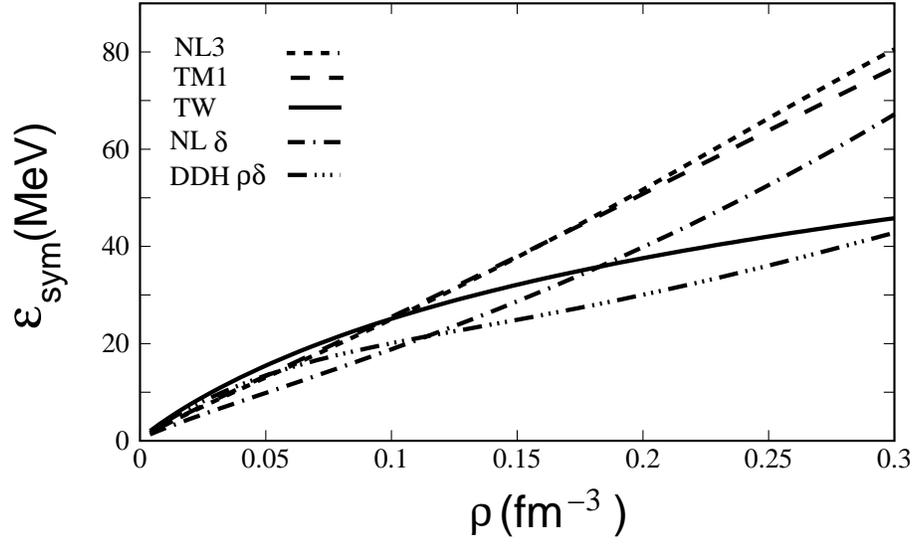}
\end{center}
\caption{Symmetry energy results for the NL3, TM1,
TW, NL$\delta$  and DDH$\rho \delta$  parameter sets.}
\label{fig2}
\end{figure}

In fig.\ref{fig2} we show the symmetry energy for the different models used in 
this work, calculated at $T=0$ and $y_p=0.5$. We can see that at 
subsaturation densities where the instability regions occur 
(0 up to 0.1 fm$^{-3}$), NL3 and TM1 parametrizations give very similar 
behaviours. The NL$\delta$ parametrization describes saturation
  at 0.16 fm$^{-3}$ and this fact affects the symmetry energy at
  low densities, in particular it is the model with the lowest
  symmetry energy in this range of densities (see Table \ref{prop}). 
The TW model presents a  different behaviour, after a faster increase at 
low densities, the symmetry energy increases much slower than most of the other
models at larger densities.  A very low value of the symmetry energy 
was obtained with  the DDH$\rho \delta$ parametrization. It amounts to 
25 MeV, in contrast with the value mentioned in \cite{gaitanos} (33.4 MeV). 
Nevertheless, in the same reference \cite{gaitanos}, the authors show a curve
in fig. 2, which confirms our number. 
The density dependent hadronic
models show  a softer symmetry energy at the densities shown.  At
higher densities the models containing the $\delta$-meson are expected
to get a harder behaviour due to relativistic effects, namely the
$\delta$ contribution goes to zero and only the repulsive
contribution from the $\rho$-meson remains \cite{lgbct}.
 We point out that in present work we are only testing the low density 
region of the symmetry energy, namely our discussion is concerned with  the behaviour 
of nuclear matter at densities below 0.7$\rho_0$, where $\rho_0$ is the 
saturation density.  In this region the quantities which better distinguish 
the different models are the slope $L=3\rho_0
\left.\left(\frac{\partial {\cal E}_{sym}}{\partial\rho}\right)\right|_{\rho=\rho_0}$ and
curvature $K_{sym}=9\rho_0^2 \left.\left(\frac{\partial^2
     {\cal E}_{sym}}{\partial\rho^2}\right)\right|_{\rho=\rho_0}$ of the 
symmetry energy \cite{lkw97} given in Table \ref{prop}.
The  TW model is the only one with a negative curvature at saturation
density. In fact,  the DDH$\rho \delta$ also has a negative curvature
at densites below 0.13 fm$^{-3}$ but at saturation the curvature has already 
a  positive value.

\begin{figure}
\begin{center}
\includegraphics[width=12.0cm,angle=0]{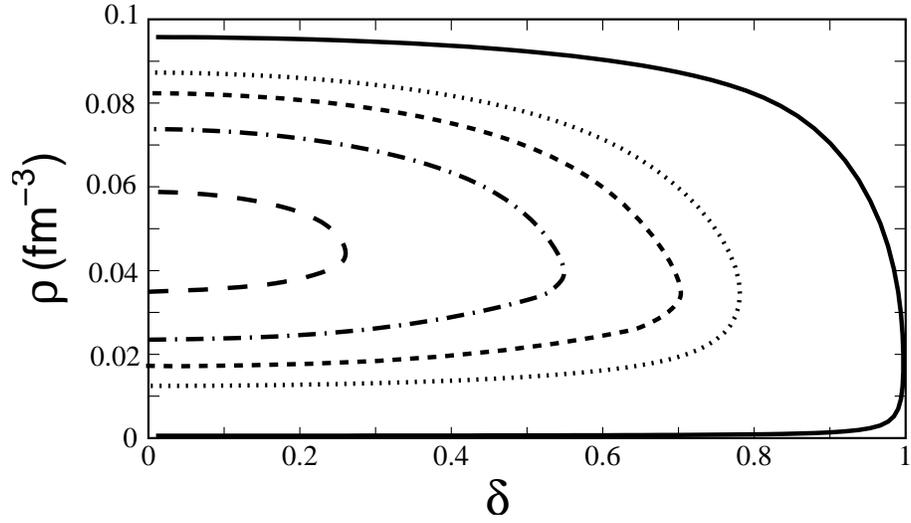}
\end{center}
\caption{Spinodal regions obtained with the NL3 parametrization and 
different temperatures. The instability regions lie inside the curves.
The curves are drawn for T=0 (solid line), 8 (dotted line), 10 (short-dashed 
line), 12 (dash-dotted line) and 14 (long dashed line) MeV.}
\label{fig3}
\end{figure}

In fig.\ref{fig3} we plot the spinodals for the NL3 parameter set and different
temperatures, varying from T=0 up to T=14 MeV, already close to the critical
temperature, $T_{NL3,c}\sim 15$ MeV. All the other models present a similar 
behaviour, namely the larger the temperature 
the smaller the instability region. These same conclusions have already been 
discussed in \cite{bao02} for Skyrme type phenomenologial EOS for ANM. After 
a critical temperature the liquid-gas phase transition is a smooth transition.
A consequence of this behavior is the greater stability
against density flucutations  of neutron-rich systems. Also, in intermediate
energy collisions fragmentation or fractionation occurs later after the 
system has cooled down.

\begin{figure}
\begin{center}
\includegraphics[width=10.cm,angle=0]{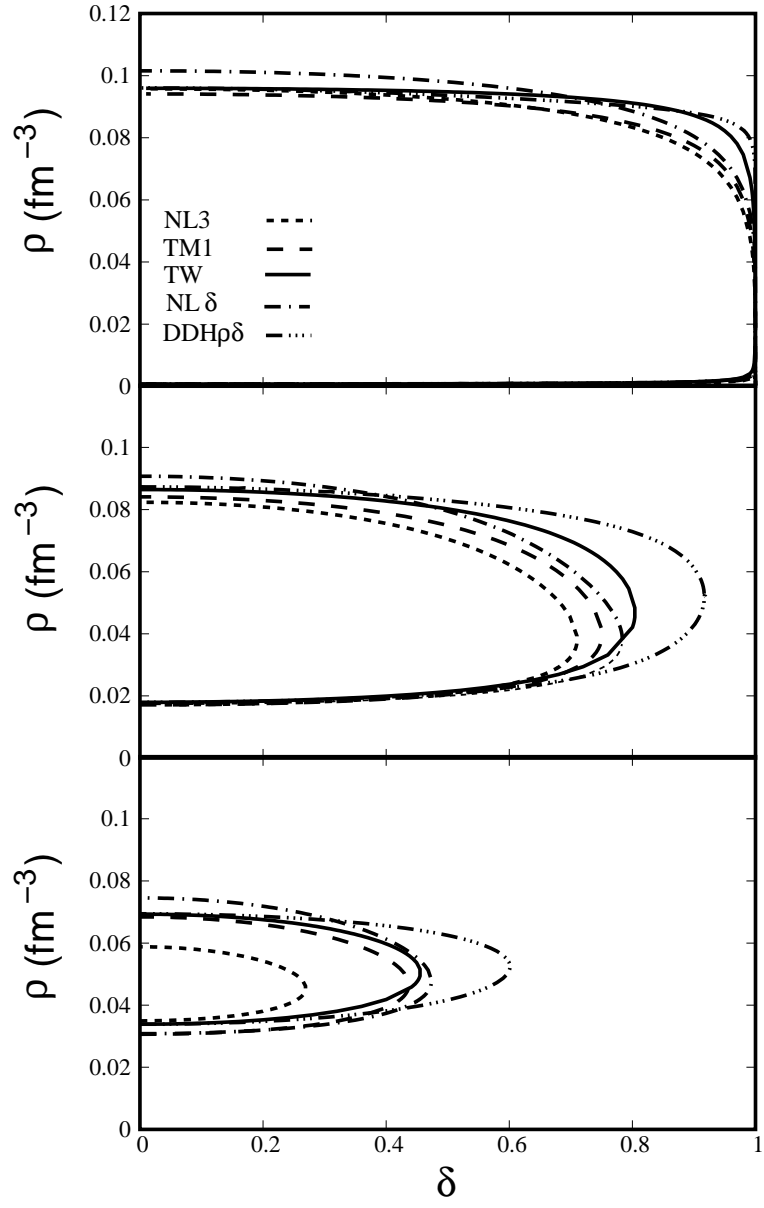} \\
\end{center}
\caption{Spinodal regions for different parameter sets and $T=0$ (upper
panel), $T=10$ MeV (middle panel) and $T=14$ MeV (lower panel).}
\label{fig4}
\end{figure}

In fig. \ref{fig4} we display the spinodals for different
parameter sets obtained with $T=$0, 10, 14 MeV.
For $T=0$ MeV the differences are not significant, occuring at the
higher density branch, $\sim 0.1$ fm$^{-3}$. In particular, in the
NL$\delta$ parametrization the boundary lies at a larger density
than in the other models for $\delta<0.5$.
 The  spinodals for the density
dependent hadronic models are decreasing  less with the
asymmetry parameter $\delta$ than the others.
At finite temperature the differences are larger. These differences
occur again in the larger density  branch, and it is again the NL$\delta$
which presents a boundary at larger densities in the small asymmetry
region followed by the TW and the DDH$\rho \delta$ models. In the large asymmetry region we are
already testing both the critical temperatures and  asymmetries of the
models and the differences between the models are larger.
For the DDH$\rho \delta$ model, as temperature increases, the
instability boundaries extend to higher asymmetries  as compared with
all the  other models.
This is possibly due to a lower symmetry energy. The opposite occurs
with the NL3 parametrization which has the larger symmetry energy.
We conclude that the information obtained from the spinodal decomposition 
 sensitive to the underlying model comes from the phase space 
close to the spinodal boundary at large isospin asymmetry.

In conclusion, we have compared the spinodal boundary as a function of 
density and isospin asymmetry, at several temperatures, for different 
relativistic models. We have considered both quantum hadrodynamical approaches 
with non-linear terms (NL3, TM1, NL$\delta$) and  density dependent hadronic
models with density dependent coupling parameters (TW, DDH$\rho\delta$). The 
largest differences between the models occur at finite temperature and are 
more clearly shown in the high isospin asymmetry region. Only physical 
quantities that explore this region in phase space, as for example  the 
neutron-proton differential flow suggested in \cite{li00}, will bring 
possible information about the EOS of ANM.  Another physical system that 
could be sensitive to present results is a protoneutron star in the process 
of stellar colapse. The physical structure of the matter, namely a smaller or 
larger non-uniform nuclear matter region, may affect properties such as the 
neutrino scattering processes \cite{rpw83}.

Another  point to be investigated is the role of the 
Coulomb interaction and finite size effects in the instabilities of  ANM. We have already shown that the electromagnetic
force cannot be neglected in nucleation processes, where droplets of lower
asymmetry are formed in a very asymmetric gas background 
either at $T=0$ or finite temperatures \cite{mp1,mp2}.  
The authors of ref. \cite{lm01} have included the Coulomb  interaction
and surface tension in their calculation of the binodals in a simplified
way and they have shown that the minimum pressure for a given temperature do 
not occur for symmetric nuclear matter and that surface effects lower
the binodal pressure.  It has also been shown that the Coulomb
interaction affects  the 
growth of instabilities 
\cite{matera}.  A natural extension of the 
present work would be the inclusion of the Coulomb interaction in the same 
spirit as done in \cite{lm01} and 
consequent construction of the binodal sections and spinodal regions for 
different temperatures.
 We expect, however, that  the overall effect of the inclusion 
of the Coulomb interaction will be the same for all the models and the main 
conclusions of the present work will not change.

\section{Appendix:  Some formulae}
In what follows $T$ is considered fixed. We give the main expressions 
needed to obtain the conditions which define the boundaries of the instability
 regions. We first consider the derivatives of the chemical 
potential with respect to $\rho$, for a given $\delta$. For this purpose we 
need the derivative of the effective chemical potential that reads:

$$\left(\frac{\partial \nu_{i}}{\partial\rho}\right)_{\delta}=\frac{\partial \nu_{i}}{\partial\eta_i}\frac{\partial
  \eta_{i}}{\partial\rho}+\frac{1}{\beta}\sum_k\eta_i^k \frac{db_k}{dM^*}\frac{\partial  M^*}{\partial\rho}$$

where
$$
\frac{\partial \nu_{i}}{\partial\eta_i}=\frac{1}{\beta}\left[\frac{1}{\eta_i}+
\sum_k \,k \, b_k\, \eta^{k-1}_i\right]\,,$$

and
$$
\frac{\partial
  \eta_{i}}{\partial \rho}=\frac{\eta_{i}}{\rho}-\frac{\beta^4  \, \pi^2 \rho_{i}}{2}\left[\frac{4K_2(x)-x\,( K_1(x)+K_3(x))}
{x^3\, K_2^2(x)}\right] \,\frac{\partial  M^*}{\partial\rho}\,,
\quad x=\beta\, M^* .$$

So, in the NLWM model, one has
$$\left(\frac{\partial \mu_{i}}{\partial\rho}\right)_{\delta}=
\left(\frac{\partial \nu_{i}}{\partial\rho}\right)_{\delta}+
\frac{g_v^2}{m_v^2}-\tau_{i3} \frac{g_\rho^2}{4m_\rho^2}\delta.$$

Instead, for the TW model, the previous equation reads
$$
\left(\frac{\partial \mu_{i}}{\partial\rho}\right)_{\delta}=\left(\frac{\partial
  \nu_{i}}{\partial\rho}\right)_{\delta}+\frac{\partial}{\partial\rho}\left(\frac{\Gamma_v^2}{m_v^2}\rho\right)-\tau_{i3}\frac{\partial}{\partial\rho}\left(\frac{\Gamma_\rho^2}{4\,m_\rho^2}\delta\,\rho\right)+\frac{\partial\Sigma^R_0}{\partial\rho},$$

with
$$\Sigma_0^R=\frac{\Gamma_v}{m_v^2} \frac{\partial\Gamma_v}{\partial\rho}\rho^2+
\frac{\Gamma_\rho}{4m_\rho^2} \frac{\partial\Gamma_\rho}{\partial\rho}\rho^2\delta^2
-\frac{m_s^2}{\Gamma_s^3}\frac{\partial\Gamma_s}{\partial\rho}\left(M-M^*\right)^2,
$$

$$M-M^*=\left(\frac{\Gamma_s}{m_s}\right)^2\rho_s.$$
\vskip0.3cm

To evaluate the expressions involved in (\ref{dmu})
we obtain the derivative of the chemical potential with respect to $\delta$
for a fixed $\rho$
$$\left(\frac{\partial \nu_{i}}{\partial\delta}\right)_{\rho}=\frac{\partial \nu_{i}}{\partial\eta_i}\frac{\partial
  \eta_{i}}{\partial\delta}+\frac{1}{\beta}\sum_k\eta_i^k \frac{db_k}{dM^*}\frac{\partial  M^*}{\partial\delta},$$
where
$$
\frac{\partial \eta_i}{\partial
    \delta}=
-\frac{\tau_{3i}\rho}{2\, \gamma Q(x)}-\frac{\beta^4  \, \pi^2 \rho_{i}}{2}\left[\frac{4K_2(x)-x\,( K_1(x)+K_3(x))}
{x^3\, K_2^2(x)}\right] \,\frac{\partial  M^*}{\partial\delta}\,.$$

Both the derivatives $\frac{\partial  M^*}{\partial\rho}$ and 
$\frac{\partial  M^*}{\partial\delta}$ are calculated numerically.

The expressions for
$\left(\frac{\partial\mu_i}{\partial\delta}\right)_{\rho}$ become 
$$
\left(\frac{\partial\mu_i}{\partial\delta}\right)_\rho= \left(\frac{\partial 
\nu_i}{\partial \delta}\right)_\rho - \tau_{3i}\left(\frac{g_\rho}
{2m_\rho}\right)^2 \rho$$
in the NLWM, and
\begin{eqnarray*}
\left(\frac{\partial\mu_i}{\partial\delta}\right)_\rho&=&\left(\frac{\partial \nu_i}{\partial
    \delta}\right)_\rho
- \tau_{3i}\left(\frac{\Gamma_\rho}{2m_\rho}\right)^2
\rho+ \frac{\Gamma_\rho}{2\, m_\rho^2} \frac{\partial \Gamma_\rho}{\partial \rho}\rho^2\delta+\frac{\partial\Sigma^R_0}{\partial\delta},
\end{eqnarray*}
in the TW model.
  
Some derivatives of the coefficients $b_l$ are also listed:
$$\frac{d\, b_1}{d M^*}=\beta\frac{d\, S_2}{d x},$$
$$\frac{d\, b_2}{d M^*}=\beta\frac{d\, b_2}{dx}=\beta\left(3 S_2\frac{d\,
  S_2}{d x}-\frac{d\, S_3}{d x}\right)\, ,$$
with
$$\frac{d\, S_n(x)}{d x}=\frac{1}{2\, n K_2^2(x)}\left[K_2(nx)\left(K_1(x)+K_3(x)\right)-n\,
  K_2(x)\left(K_1(nx)+K_3(nx)\right)\right]\,.$$

\section*{ACKNOWLEDGMENTS}
This work was partially supported by CNPq (Brazil), 
CAPES(Brazil)/GRICES (Portugal) under project 100/03 and FEDER/FCT (Portugal) 
under the project POCTI /35308/ FIS/ 2000.

\vspace{0.5cm}
\begin{table}[h]
\caption{ Nuclear matter properties.}
\label{prop}
\vspace{0.5cm}
\begin{center}
\begin{tabular}{|c|c|c|c|c|c|c|c|c|c|c|}
\hline
&  NL3 \cite{nl3}  &  TM1 \cite{tm1} & TW \cite{TW} & NL$\delta$ \cite{lgbct}
& DDH$\rho \delta$ \cite{gaitanos}\\
\hline
$B/A$ (MeV) & 16.3 & 16.3 & 16.3 & 16.0 & 16.3\\
\hline
$\rho_0$ (fm$^{-3}$) & 0.148 & 0.145 & 0.153 & 0.160 & 0.153\\
\hline
$K$ (MeV) & 272 & 281 & 240 & 240 & 240 \\
\hline
${\cal E}_{sym.}$ (MeV)  & 37.4 & 36.9 & 32.0 & 30.5 & 25.1 \\
\hline
$M^*/M$ & 0.60 & 0.63 & 0.56 & 0.60 & 0.56\\
\hline
$L$ (MeV)&123. &  117. &   55.  & 101. &44.\\    
\hline
$K_{sym}$ (MeV)&108. &36 & -124.& 112.&45.\\
\hline
\end{tabular}
\end{center}
\end{table}

\begin{table}[h]
\caption{Parameters of the $DDH\rho\delta$ model}
\label{para}
\vspace{0.5cm}
\begin{center}
\begin{tabular}{lccccc}
\hline
i&$\Gamma_i$&$a_i$&$b_i$&$c_i$&$d_i$\\
\hline
$\rho$& 5.8635&0.095268&2.171&0.05336&17.8431\\
\hline
$\delta$&7.58963&0.01984&3.4732 &-0.0908&-9.811\\
\hline
\end{tabular}
\end{center}
\end{table}

\newpage

\end{document}